\documentclass[aps,amsmath,amssymb,preprintnumbers,groupedaddress,twocolumn]{revtex4}
\usepackage{graphicx}

\newcommand{\Z}{\mathbb{Z}}

\newcommand{\drawsquare}[2]{\hbox{%
\rule{#2pt}{#1pt}\hskip-#2pt
\rule{#1pt}{#2pt}\hskip-#1pt
\rule[#1pt]{#1pt}{#2pt}}\rule[#1pt]{#2pt}{#2pt}\hskip-#2pt
\rule{#2pt}{#1pt}}

\newcommand{\Ysymm}{\raisebox{-.5pt}{\drawsquare{6.5}{0.4}}\hskip-0.4pt%
        \raisebox{-.5pt}{\drawsquare{6.5}{0.4}}}
\newcommand{\Yasymm}{\raisebox{-3.5pt}{\drawsquare{6.5}{0.4}}\hskip-6.9pt%
        \raisebox{3pt}{\drawsquare{6.5}{0.4}}}

\newcommand{\nc}{\newcommand}
\nc{\beq}{\begin{equation}} \nc{\eeq}{\end{equation}} \nc{\bea}{\begin{eqnarray}}
\nc{\eea}{\end{eqnarray}}

\def\gsim{\mathrel{\rlap{\lower4pt\hbox{\hskip1pt$\sim$}}
    \raise1pt\hbox{$>$}}}       

\def\K3{{\bf K3}}

\def\n2d{\cN_{V^*}^{\otimes 2}}

\def\cN{{\mathcal N}}

\begin{document}

\preprint{MIFP-09-29, ACT-07-09}

\title{Quark Masses from Gaugino Condensation in String Theories}

\author{Christos Kokorelis $^{1 }$}\email{kokorelis@inp.demokritos.gr}
\author{Van E. Mayes $^{2 }$}\email{eric@physics.tamu.edu}
\author{Dimitri V. Nanopoulos $^{2,3,4}$}\email{dimitri@physics.tamu.edu}
\affiliation{$^{1}$ Institute of Nuclear Physics, N.C.S.R. Demokritos, 15310, Athens, Greece\\
$^{2}$ George P. and Cynthia W. Mitchell Institute for Fundamental Physics, Texas A$\&$M University, College Station, TX 77843, USA\\
$^{3}$ Astroparticle Physics Group, Houston Advanced Research Center (HARC), Mitchell Campus, Woodlands, TX 77381, USA\\
$^{4 }$ Academy of Athens, Division of Natural Sciences, 28 Panepistimiou Avenue, Athens 10679, Greece}

\begin{abstract}
\noindent
We present a mechanism able to generate the perturbatively absent
up/down $\langle {\bf 10} \cdot {\bf 10} \cdot {\bf 5}^H \rangle$ quark Yukawa couplings of SU(5)/flipped SU(5) GUTS in Type II orientifold compactifications with D-branes.
 The mechanism works when there are Sp(N) gauge groups involved.
The ${\bf {\bar 5}}$'s get charged under the  Sp(N) gauge groups and the
generation of quark masses proceeds via the generation of the fermionic Sp(N) singlet
condensate $\langle {\bf {\bar 5} \cdot {\bar 5} \cdot {\bar 5} \cdot {\bar 5}}\rangle $ in the term
$( 1/{M_s^5}) {\bf 10} \cdot {\bf 10} \cdot \langle {\bf {\bar 5} \cdot {\bar 5} \cdot {\bar 5} \cdot {\bar 5}} \rangle $. Also non-chiral states charged under Sp gauge groups may become constrained by the requirement of Sp's becoming strongly coupled.

\end{abstract}


\maketitle

\date{today}

\bigskip

\section{INTRODUCTION}

Non-perturbative mechanisms are very important in modern particle theories as they are used to solve various problems.
For example, QCD instantons \cite{'t Hooft:1976fv} are a famous solution in QCD for solving the $\eta'$-meson mass problem \cite{Witten:1978bc} and SU(2) instantons in generating baryon number violating interactions.
On the other hand in string compactifications worldsheet instantons have been associated with genus zero instantons that generate corrections to the spacetime superpotential \cite{Dine:1986zy}.
 Moreover gaugino condensation has been connected to the formation of
 hidden matter condensates \cite{Lust:1990zi} and supersymmetry breaking.
 Also in the context of intersecting brane worlds
  \cite{Blumenhagen:2005mu,Blumenhagen:2006ci} hidden sector confinement has
  been utilized to decouple hidden sector states at high energies \cite{Cvetic:2002qa}.
  On the other hand as intersecting brane models (IBM) \cite{Blumenhagen:2005mu,Blumenhagen:2006ci} fix some of
   the moduli either via supersymmetry conditions in intersections or fluxes,
    it is quite desirable to find a flipped SU(5) (FSU(5)) IBM string model which could
    mimic the phenomenological success of fermionic FSU(5) \cite{Antoniadis:1989zy}.

Quite recently, in the context of type II compactifications, D2-brane (E2) instantons wrapping
special Lagrangian three-cycles of the internal Calabi-Yau \cite{Blumenhagen:2009qh}
can generate perturbatively absent couplings like Majorana terms for right handed neutrinos, $\mu$-terms in the MSSM \cite{Blumenhagen:2009qh,Ibanez:2008my,Cvetic:2009yh} or solving via instantons \cite{Blumenhagen:2007zk} the long standing problem of the missing \cite{Bachas:1995ik,Blumenhagen:2001te,Ellis:2002ci,Axenides:2003hs,Kokorelis:2004gb,
Chen:2006ip,Bachas:2008jv,Chen:2005aba,Chen:2005cf}
up/down quark mass ${\bf 10} \ {\bf 10} \ {{\bf 5}^H}$
coupling of SU(5)/flipped SU(5) GUTS (from orientifolds of type II compactifications). The latter problem have been also been shown to be solvable by a combination of condensates and Higgs fields by a higher order coupling \cite{Kokorelis:2008ce}.

 In this work we describe a new way of generating the missing ${\bf 10} \ {\bf 10} \ {{\bf 5}^H}$ in orientifolds of type IIA compactifications by utilizing the formation of strongly coupled condensates extending the work of \cite{Kokorelis:2008ce}. In the first part of the this work, we discuss the generation of down-quark masses in flipped SU(5)(FSU(5)). In the second part we generate the up-quark masses in SU(5) GUTS via the same mechanism.


\section{The missing Yukawa coupling term}
\subsection{FLIPPED SU(5) MODELS}
Grand unified theories giving rise to SU(5) type gauge group structures \cite{Bachas:1995ik,Blumenhagen:2001te,Ellis:2002ci,Axenides:2003hs,
Kokorelis:2004gb,Chen:2005aba,Chen:2005cf,Chen:2006ip,Bachas:2008jv}
arising from 4D compactifications of type II theories on N=1 supersymmetric intersecting brane orientifolds are missing the Yukawa coupling ${\bf 10} \  {\bf 10} \  {\bar 5}^H$, which is responsible for generating masses for the down quarks
\cite{Ellis:2002ci,Axenides:2003hs,Kokorelis:2004gb,Chen:2005aba,
Chen:2005cf}).
 In an SU(5) GUT the same coupling gives rise to a missing up-quark mass (e.g. see
\cite{Blumenhagen:2001te,Axenides:2003hs,Kokorelis:2004gb,
Chen:2006ip,Bachas:2008jv}).

In \cite{Kokorelis:2008ce}, a candidate solution for a perturbative term to the mass of the up-quarks in SU(5) GUTs was suggested where string models with gauge groups in the form  $SU(5) \times U(1)^n$ with $n \in Z$ were used. We have argued that the form of the higher dimensional coupling that gives a mass to the missing $10 \ 10 \ 5^H$ SU(5) Yukawa's appears from an operator in the form
$ \frac{1}{M_s^7}\ {\bf 10} \ {\bf 10} \ {\bf \overline{5}} \ {\bf \overline{5}} \ {\bf \overline{5}} \ {\bf \overline{5}} \  1^H \ 1^H$
 where the $1^H$ gauge singlets are used to cancel the surplus bifundamental charge. In this case, the up-quark mass may come from the condensation of ${\bar 5}$ fermions
Lets us consider the field theory operator $\ {\bf 10} \ {\bf 10} \ {\bf \overline{5}} \ {\bf \overline{5}} \ {\bf \overline{5}} \ {\bf \overline{5}} $ and
 rewriting this term in SU(5) notation as
\beq
\frac{1}{M_s^5}\epsilon^{ijklm} \ {\bf 10}^{ij} \ {\bf 10}^{kl } \   \epsilon^{m \phi \chi \psi \omega} \ {\bf {\bar 5}}^{\phi} \ {\bf {\bar 5}}^{\chi} \ {\bf {\bar 5}}^{\psi} \ {\bf {\bar 5}}^{\omega},
\label{ki2}
\eeq
we conclude that in components eq. (\ref{ki2}) reads
\beq
{\bf 10}^{12} \ {\bf 10}^{34} \  {\bf {\bar 5}}^{1} \ {\bf {\bar 5}}^{2} \ {\bf {\bar 5}}^{3} \ {\bf {\bar 5}}^{4}
\equiv (u u^c_L ) ( d^c_L d^c_L d^c_L e_L ).
\label{contrib1}
\eeq
In \cite{Kokorelis:2008ce} it was argued that, in the context of field theory
(\ref{contrib1}), can only generate a mass as long as the fermions of the Higgs-like ${\bar 5}$-plets condense.
In this work, we will suggest that an alternative explanation for the appearance for such a fermion condensate exists, within in the context of SU(5) models constructed on Type IIA orientifolds with intersecting D6-branes where the SU(5) gauge symmetry is extended by Sp(N) gauge groups.
The binding mechanism is easy to understand in terms of the gauge forces acting on
the fermion condensate as the ${\bf \overline{5}}$ condense and are charged under Sp(2). The order of the condensate is
$\langle (\overline{5}, 1, 1, 2)^4 \rangle = \Lambda^6$, where $\Lambda$ is the scale the fermion condensate forms.
The mass of the up-quarks is determined from (\ref{contrib1}) and for the present model it reads
\beq
M_{quarks}^U \approx \frac{\Lambda^6}{M_s^5}.
\label{rel1}
\eeq

Let us also consider the flipped SU(5) model \cite{Chen:2006ip} FSU(5)-IV on Type IIA T6/($Z2 \times Z2$) orientifold that also includes metric, NSNS, and RR fluxes \cite{Villadoro:2005cu}, \cite{Camara:2005dc}.
In this flipped SU(5) model the complete gauge symmetry is $U(5) \times U(1)^{10} \times
USp(10)^2 \times USp(8) \times USp(2)$.

\begin{table}[h]
\renewcommand{\arraystretch}{.1}
\begin{center}
\footnotesize
\begin{tabular}{|c|c||c|c|c|c|} \hline

stk & $N$ & ($n_1$,$l_1$)($n_2$,$l_2$) ($n_3$,$l_3$) & A & S
& $O64$    \\ \hline \hline

$a$ & 10 & ( 1, 3)( 1, 1)( 0,-1) & 2 & -2 & -1\\ \hline

$b$ & 2 & ( 1,-3)( 0,-1)(-1, 1) & -2 & 2 & 0\\ \hline

$c$ & 2 & ( 1, 3)(-1, 1)(-1, 0) & 2 & -2 &0 \\  \hline

$d$ & 2 & ( 1, 1)(-1,-3)( 0, 1) & 2 & -2  &-1 \\  \hline

$e$ & 2 & (-5, 9)(-5,-3)( 1, 0) & -30 & 30 & 0\\  \hline

$f$ & 2 & ( 2, 0)(-1, 3)(-1,-3) & 0 & 0   & 6 \\ \hline

$g$ & 2 & ( 1,-9)(-1, 0)(-1,-3) & -6 & 6 & 3 \\ \hline

$h$ & 2 & ( 1,-7)( 0, 1)( 7,-3) & 0 & 0 & 0\\ \hline

$i$ & 2 & ( 0, 2)( 4,-3)( 3,-4) & 0 & 0 & 0\\ \hline

$j$ & 2 & ( 1,-3)(-1, 0)(-1,-1) & -2 & 2 &1\\ \hline

$k$ & 2 & ( 0, 2)(-3,-1)( 1, 3) & 0 & 0 &0 \\ \hline

$O6^1$ & 10 & ( 2, 0)( 1, 0)( 1, 0) & - & - &-\\ \hline

$O6^2$ & 10 & ( 2, 0)( 0,-1)( 0, 1) & - & -& -\\ \hline

$O6^3$ & 8 & ( 0,-2)( 1, 0)( 0, 1) & - & - & -\\ \hline

$O6^4$ & 2 & ( 0,-2)( 0, 1)( 1, 0) & - & - & -\\  \hline
\end{tabular}
\caption{D6-brane configurations and intersection numbers (Part 1)
for the Model FSU(5)-IV on Type IIA $\mathbf{T^6/(\Z_2\times
\Z_2)}$ orientifold with fluxes. The
complex structure parameters are $\chi_1 =2/3, \;\; \chi_2=1$, and
$\chi_3 =1$. We have listed only the intersections of the branes with the Sp(2) generating brane.}
\label{FSU5-4}
\end{center}
\end{table}

We observe that only the Yukawa mass coupling term
giving a mass to the up quarks is allowed (there is no obvious mass term for the leptons as well)
\beq
\langle {\bf 10}_{(2, 0^{10})}  {\bf \overline{5}}_{(-1, 1, 0^9)}
{\bf \overline{5}}^H_{(-1,-1, 0^9)} \rangle, \  \langle {\bf \overline{5}}_{(-1, 1, 0^9)}
{\bf 1}_{(0, 2, 0^9)}  {\bf 5}^H_{(1, 1, 0^9)} \rangle
\eeq
We used for convenience the notation $k^n \equiv (k, k, ..., k)$ with k repeated n times.
Also as there is no mass term for the down-quarks of the form
$
{\bf 10}_{(2, 0^{10})} \cdot {\bf 10}_{(2, 0^{10})} \cdot {5}^H_{(1, 1, 0^9)}$
as this is not allowed by U(1)$_{a}$,U(1)$_b$ charge conservation. This is the usual problem of any SU(5)/flipped SU(5) GUT from any type II
orientifold compactifications e.g.\cite{Blumenhagen:2001te,Ellis:2002ci,Axenides:2003hs,Kokorelis:2004gb,Chen:2005aba,
Chen:2005cf}.
 The down-quark masses may be generated by the expression
 \beq
\frac{1}{M_s^5} \left((10, 1^{10};1^3,1)_{(2, 0^{10})}\right)^2 \cdot \left((\overline{5}, 1^{10};1^3, 2)_{(-1, 0^{10})}^H \right)^4 \ .
\label{det}
\eeq
In order to show that the ${\bar 5}'$s condense we will adjoint split the gauge group U(5)$\rightarrow$ U(3) x U(2), so that the U(5)-D6-brane splits into a and a1 branes, each one associated with the U(3), U(2) brane stacks. The anomaly free hypercharge corresponding to the decomposition of table (\ref{tableena1}) is
\bea
Y=\frac{1}{6}U(1)_a -\frac{1}{2}U(1)_b + \frac{1}{2}U(1)_c + \frac{1}{2}U(1)_d &\nonumber\\ -\frac{1}{2}U(1)_j +\frac{1}{2}U(1)_e +\frac{1}{2}U(1)_f-\frac{1}{2}U(1)_g \nonumber\\ -\frac{1}{2}U(1)_h +\frac{1}{2}U(1)_i -\frac{1}{2}U(1)_k
\eea
\begin{table}[ht]
\centering
\begin{tabular}{|c|c||c|c||c|}
\hline
    & \multicolumn{4}{c|}{$SU(3)\times SU(2)_{a_1} \times U(1)_b \cdots \times USp(2)$} \\
\hline\hline Intersection & \rm{Multiplicity} & & &Y \\
\hline
 a,2 & $1$ && $({\bar 3}, 1, 1^{10}; 1, 1, 1, 2)_{(-1, 0, 0^{10})}$ & $-\frac{1}{6}$\\
 a$_1$,2& $1$  && $(1, 2, 1^{10}; 1, 1, 1, 2)_{(0, -1, 0^{10})}$& $0$\\
 b,2 & $B'$ && $(1^{12};1, 1, 1, 2)_{(0, 0, 1, 0^9)}$& $-\frac{1}{2}$\\
 b,2 & $B'$ && $(1^{12};1, 1, 1, 2)_{(0, 0, -1, 0^9)}$ & $\frac{1}{2}$ \\
 c,2 & $C'$ && $(1^{12}; 1, 1, 1, 2)_{(0, 0, 0, 1, 0^8)}$& $\frac{1}{2}$\\
 c,2 & $C'$ && $(1^{12}; 1, 1, 1, 2)_{(0, 0, 0, -1, 0^8)}$& $-\frac{1}{2}$\\
 d,2 & $1$  && $(1^{12}; 1, 1, 1, 2)_{(0^4, -1, 0^{7})}$ & $\frac{1}{6}$\\
 e,2 & $E'$ && $(1^{12}; 1, 1, 1, 2)_{(0^5, 1, 0^6)}$           &$\frac{1}{2}$\\
 e,2 & $E'$ && $(1^{12}; 1, 1, 1, 2)_{(0^5, -1, 0^6)}$           &$-\frac{1}{2}$\\
 f,2 & $6$ && $(1^{12}; 1, 1, 1, 2)_{(0^6, 1, 0^5)}$  &$\frac{1}{2}$\\
g,2   & $3$ && $(1^{12}; 1, 1, 1, 2)_{(0^7, 1, 0^4)}$ & $-\frac{1}{2}$\\
h,2   & $H'$ &&  $(1^{12};  1, 1, 1, 2)_{(0^8, 1, 0^3)}$& $-\frac{1}{2}$\\
h,2   & $H'$ && $(1^{12}; 1, 1, 1, 2)_{(0^8, -1, 0^3)}$ & $\frac{1}{2}$ \\
i,2   & $I'$ &&  $(1^{12};  1, 1, 1, 2)_{(0^9, 1, 0^2)}$& $\frac{1}{2}$\\
i,2   & $I'$ && $(1^{12}; 1, 1, 1, 2)_{(0^9, -1, 0^2)}$ & $-\frac{1}{2}$ \\
j,2   & $1$ && $(1^{12}; 1, 1, 1, 2)_{(0^{10}, 1, 0)}$ & $-\frac{1}{2}$ \\
k,2   & $K'$ && $(1^{12}; 1, 1, 1, 2)_{(0^{11}, 1)}$ & $-\frac{1}{2}$ \\
k,2   & $K'$ && $(1^{12}; 1, 1, 1, 2)_{(0^{11}, -1)}$ & $\frac{1}{2}$ \\
\hline
\end{tabular}
\caption{N=1 multiplets charged under Sp(2) in FSU(5).
 The gauge group is $SU(3)\times SU(2)_{a_1} \times U(1)_b \times U(1)_c \times U(1)_d \times U(1)_e \times U(1)_f \times U(1)_g \times U(1)_h \times U(1)_i \times U(1)_j \times U(1)_k
 \times USp(10)^2 \times USp(8)
    \times USp(2)$.}
\label{tableena1}
\end{table}
We have parametrized the number of non-chiral states in intersections where branes are parallel in at last one tori by B$'$, C$'$, E$'$, H$'$, I$'$, K$'$.
The $\beta$-function for Sp(2) is
$b^{Sp(2)}= 8+B'+C'+E'+H'+I'+K'-6$ and may become negative as long the number of non-chiral states is either 0 or if they decouple at a typical mass of order $M_s$, higher than the condensation scale $\Lambda_{cond}$.
We may also need to give a mass $\Lambda_{cond}$ to the multiplets from the intersections (a,2), (a$_1$,2), (f,2), (g,2).
The leptons may get a mass from
\beq
\frac{1}{M_s^{6}}{\bf \overline{5}}_{(-1, 1, 0^9)}
{\bf 1}_{(0, 2, 0^9)}  {\bf 5}^H_{(1, 1, 0^9)} \langle (1^{12};1,1,1,2)_{(0, -1, 0^{10})} \rangle^4
\eeq
where the states $\langle \cdots \rangle$ from the intersection (a$_1$,2) may condense and decouple. In Table \ref{tria1}, using eq.(\ref{rel1}), we exhibit the condensation scale against the string scale given the phenomenological down-quark masses; $3.5< m_d < 6$ MeV, $M_s \approx 105$ MeV, $M_b \approx 4.20$ GeV \cite{parti}.
\begin{table}[ht]
\centering
\begin{tabular}{|c|c||c|c|c|}
\hline
GeV & & \rm{$M_S = 10^{16} $}&\rm{$M_S = 10^{17}$} & \rm{$M_S = 10^{18}$}  \\\hline
\hline Quark & \rm{$M_{quark}$} & & & \\
\hline
   d&  $5 \cdot 10^{-3}$ &$8.9 \cdot 10^{12}$ & $3.25 \cdot 10^{14}$ & $4.1 \cdot 10^{14}$\\\hline
  s&  $105 \cdot 10^{-3}$ & $1.5 \cdot 10^{13}$ & $1.0 \cdot 10^{14}$& $6.9 \cdot 10^{14}$\\\hline
 b &  $4.2$ &$2.7 \cdot 10^{13}$ &  $1.9 \cdot 10^{14}$ & $ 1.3 \cdot 10^{15}$\\
\hline
\end{tabular}
\caption{$\Lambda_{cond}$ vs down-quark masses {\em using eq.(\ref{rel1})}.}
\label{tria1}
\end{table}
Thus a generic FSU(5) string model may therefore describe the masses of the down-quarks as long as its condensation scale is in the range
\beq
10^{12} \ GeV < \ \Lambda^{cond}_{FSU(5)} \ < \ 10^{15} \ GeV
\label{eqn10}
\eeq
 If we consider the running gauge Sp(2) coupling following \cite{Cvetic:2002qa} with $b_{Sp(2)}=-2$ where have set, B$'$, C$'$, E$'$, H$'$,I$'$,K$'$ to zero and the
states from intersections (a,2), (a$_1$,2), (g,2) decouple, $\Lambda_{cond}$
is determined in table (\ref{tria2}). $M_s$ can be higher than $10^{18}$ GeV since its value depends on the number of non-chiral states which we will not determine here.
\begin{table}[ht]
\centering
\begin{tabular}{|c||c|c|c|c|}
\hline
 &\rm{$M_S = 10^{16} $} & \rm{$M_S = 10^{18} $}&\rm{$M_S = 10^{19}$} & \rm{$M_S = 10^{21}$ }  \\\hline
\hline
  $\Lambda^{cond}_{run}$    & $2.6 \cdot 10^{10}$  & $2.6 \cdot 10^{12}$ & $2.6 \cdot 10^{13}$ & $2.6 \cdot 10^{14}$\\\hline
\end{tabular}
\caption{$\Lambda^{cond}_{run}$ from the running Sp(2) gauge coupling.}
\label{tria2}
\end{table}
 Values of $\Lambda^{cond}$ higher from $\Lambda^{cond}_{run}$ may be understood (from the exact string amplitude of the associated fermionic correlator of eq.(\ref{contrib1})) which involves a suppression factor lowering $\Lambda^{cond}$ to $\Lambda^{cond}_{run}$.

\section{SU(5) GUTS}

{\bf Let} us consider for example the SU(5)-type D6-brane model II.1.4. of \cite{Cvetic:2002pj}. It requires  three stacks $a$, $c$ and $d$ of D6-branes giving rise to a
$U(5)_a\times U(1)_c\times U(1)_d$ gauge symmetry intersecting at angles in IIA orientifolds of $Z_2 \times Z_2$  toroidal compactifications \cite{Cvetic:2001nr}.
 The $U(5)_a$ splits into $SU(5)_a\times U(1)_a$, where the anomalous
$U(1)_a$ gauge boson becomes massive via the generalized Green-Schwarz mechanism and
$U(1)_a$ appears as a global symmetry in
the effective action. The matter transforming as $\bf{10}$ under $SU(5)_a$ arises at intersections of
stack $a$ with its image $a'$;
the matter fields transforming as $\bf{\bar{5}}$ as well as Higgs
fields $\bf{5}_H$ and $\bf{\bar{5}}_H$ are located at intersections of stack $a$ with $c$ and $c'$ or $d$ and $d'$.
The key input in  the construction of the D-brane model is summarized in Tables \ref{wn1} and \ref{sp}. Table \ref{sp} lists the candidate Higgs fields.

\begin{table}[ht]
\begin{tabular}{|c||c|c||c|c|c|c|c|c|c|}
\hline
  & \multicolumn{9}{c|}{$U(5)\times U(1)_c \times U(1)_d \times USp(2)$} \\
    \hline\hline \rm{stack} & $N$ & $(n^1,l^1)(n^2,l^2)
(n^3,l^3)$ & $n_{\Ysymm}$ & $n_{\Yasymm}$& $c$ & $d$ & $c'$ & $d'$ & 2\\
\hline
    $a$&  10& $(0,-1)(1,4)(1,1)$ & -3 & 3  & 0 & 18 & 0 & 8 &-1\\
    $c$&   2& $(-1,3)(-1,4)(1,1)$ & -6 & -42 &- & -18  & - & -72 & -3\\
    $d$&   2& $(-1,0)(-1,2) (7,1)$ & 13  & -13 &- &- & -& -&0 \\
\hline
    2&2 & $(1,0)\times (0,1)\times (0,-2)$ &\multicolumn{7}{c|}{$7x_B=28x_A=20x_C$}\\
\hline
\end{tabular}
\caption{Wrapping numbers of D6-branes.}
\label{wn1}
\end{table}
\begin{table}[ht]
\centering
\begin{tabular}{|c|c|c|}
\hline
sector & number &  $U(5)_a \times U(1)_c \times U(1)_d \times USp(2)$   \\
\hline \hline
$(a,a')$ &  $3$ & ${ {{(10, 1, 1, 1)}}}_{(2, 0, 0)} + {(\overline{15}, 1, 1, 1)}_{(-2,0,0)} $     \\
$(a,c)$ &  $0$ & ${ (5, 1, 1, 1)_{(1, -1, 0)}} +{({\overline{5}, 1, 1, 1)}_{(-1, 1, 0)}}$    \\
$(a,c')$ & $0$ & ${(\overline{5}, 1, 1, 1)}_{(-1, -1, 0)}^H + {({5}, 1, 1, 1)}_{(1,1,0)}^H $      \\
$(a,2)$ &  $-1$ & ${ ({\overline{5}}, 1, 1, 2)}_{(-1, 0, 0)}^H$  \\
\hline
\end{tabular}
\caption{N=1 multiplet matter spectrum for $SU(5)$.
}
\label{sp}  
\end{table}
As can be seen from tables (\ref{wn1}) and (\ref{sp}), the only mass terms allowed are the ones that
are associated with the Yukawa couplings giving masses to the down quarks and leptons respectively
\beq
\langle {\bf 10}_{(2, 0, 0)}  {\bf \overline{5}}_{(-1, 1, 0)}
{\bf \overline{5}}^H_{(-1,-1, 0)} \rangle, \  \langle {\bf \overline{5}}_{(-1, 1, 0)}
{\bf 1}_{(0, -2, 0)}  {\bf 5}^H_{(1, 1, 0)} \rangle
\eeq
and
there is no mass term for the up-quarks of the form
\beq
{\bf 10}_{(2, 0, 0)} \cdot {\bf 10}_{(2, 0, 0)} \cdot {5}^H_{(1, 1, 0)}
 \eeq
as this is not allowed by U(1)$_{a}$,U(1)$_b$ charge conservation. This is the usual problem of any SU(5) GUT from any type II
orientifold compactifications e.g.\cite{Blumenhagen:2001te,Ellis:2002ci,Axenides:2003hs,Kokorelis:2004gb}.
Thus, working without loss of generality in the context of SU(5) model shown in Tables (\ref{wn1}) and (\ref{sp}), a perturbative mass term
for the up-quark masses of the form
\beq
\frac{1}{M_s^5} \left((10, 1, 1, 1)_{(2, 0, 0)}\right)^2 \cdot \left((\overline{5}, 1, 1, 2)_{(-1, 0, 0)}^H \right)^4 \ .
\label{det1}
\eeq
exists.
In order to demonstrate that the Sp(2) gauge group will become strongly coupled, we will use adjoint breaking (AB) to break the SU(5) down to the $SU(3)_a \times SU(2)_b \times U(1)_Y$, $Y=(1/6)U(1)_a +(1/2)U(1)_b -(1/2)U(1)_c+(1/2)U(1)_d$, which is equivalent to splitting the stacks on one torus. To establish notation,  $U(1)_a$, $U(1)_b$ are the U(1)'s within U(3), U(2) (of U(5)). Thus, we only present explicitly the states charged under Sp(2) in table (\ref{tableena2}).
We have parametrized the
appearance of an arbitrary number of non-chiral states becoming massless by A.
\begin{table}[ht]
\centering
\begin{tabular}{|c||c|c||c|}
\hline
     & \multicolumn{3}{c|}{$SU(3)\times SU(2)_w \times U(1)_c \times U(1)_d \times USp(2)$} \\
\hline\hline \rm{Multiplicity} & & &Y \\
\hline
    $1$ && $({\bar 3}, 1, 1, 1, 2)_{(-1, 0, 0, 0)}$ & $-\frac{1}{6}$\\
   $1$ && $(1, 2, 1, 1, 2)_{(0, -1, 0, 0)}$& $-\frac{1}{2}$\\
  $3$ &&  $(1, 1, 1, 1, 2)_{(0, 0, -1, 0 )}$& $\frac{1}{2}$\\
  $A$ && $(1, 1, 1, 1, {\bar 2})_{(0, 0, 0, 1)}$ & $\frac{1}{2}$ \\
  $A$ && $(1, 1, 1, 1, 2)_{(0, 0, 0, -1)}$& $-\frac{1}{2}$\\
\hline
\end{tabular}
\caption{N=1 multiplets charged under Sp(2).}
\label{tableena2}
\end{table}
Since $b_a^{Sp(2)} = A - 2$, the $b$-function becomes asymptotically free when the number of non-
chiral multiplets becomes either 0 or 1.

In Table \ref{tria}, utilizing the relation (\ref{rel1}),
 we show how the condensation scale depends on the string scale
 given the values of the phenomenologically determined up-quark masses
 with m$_u$ = 1.5 to 3.3 MeV; $m_c=1.27$ GeV, $m_t \approx 171$ GeV \cite{parti}.

\begin{table}[ht]
\centering
\begin{tabular}{|c|c||c|c|c|}
\hline
Quark&\rm{$GeV$} & \rm{$M_S = 10^{16} $}&\rm{$M_S = 10^{17}$} & \rm{$M_S = 10^{18}$}  \\\hline
\hline
   c&  $1.27$ &$2.2 \cdot 10^{13}$ & $1.5 \cdot 10^{14}$ & $1.0 \cdot 10^{15}$\\\hline
  u&  $2.5 \cdot 10^{-3}$ & $7.9 \cdot 10^{12}$ & $5.4 \cdot 10^{13}$& $3.7 \cdot 10^{14}$\\\hline
 t &  $171.3$ &$5.1 \cdot 10^{13}$ &  $3.4 \cdot 10^{14}$ & $ 2.3 \cdot 10^{15}$\\
\hline
\end{tabular}
\caption{$\Lambda_{cond}$  (GeV) vs up-quark masses {\em using eq.(\ref{rel1})}.}
\label{tria}
\end{table}
Thus a generic SU(5) string model may therefore describe the masses of the up-quarks as long as its condensation scale is in the range \newline
\beq
10^{12} \ GeV < \ \Lambda^{cond}_{SU(5)} \ < \ 10^{15} \ GeV
\label{eqn20}
\eeq
The range of $\Lambda^{cond}_{run}$ is what
 one gets from the strong coupling of Sp is similar to that of FSU(5)(we have set $A=0$, $\chi_C=28/20$).
\begin{table}[ht]
\centering
\begin{tabular}{|c||c|c|c|c|}
\hline
 &\rm{$M_S = 10^{16} $} & \rm{$M_S = 10^{18} $}&\rm{$M_S = 10^{19}$} & \rm{$M_S = 10^{21}$ }  \\\hline
\hline
  $\Lambda^{cond}_{run}$    & $2.4 \cdot 10^{9}$  & $2.4 \cdot 10^{11}$ & $2.4 \cdot 10^{12}$ & $2.4 \cdot 10^{14}$\\\hline
\end{tabular}
\caption{$\Lambda^{cond}_{run}$ from the running Sp(2) gauge coupling.}
\label{tria2}
\end{table}

We have presented a new perturbative mechanism to generate
the masses of the up/down quarks through fermion condensates within N=1
supersymmetric SU(5)/flipped SU(5) constructions from intersecting D6-branes. We have also demonstrated that as $\Lambda^{cond}$ appears to be in the same range as that expected from the running of Sp $\Lambda^{cond}_{run}$
for up and down quarks (eqn's (\ref{eqn10}),(\ref{eqn20})), this indicates
possible flavour independence. Also, from the $\Lambda ^6$ dependence
on the condensation scale, expected small differences
in condensate values naturally may provide us with the
fermion mass textures.
It would be also important to further study
this issue by calculating the relevant string amplitudes.
 We have also seen that a necessary condition for the existence of the fermion condensate
giving a mass to the missing quark masses, is to simultaneously demand
that the number of non-chiral states charged under the Sp(N) condensing
gauge group may be fixed so that Sp(N) may have negative $\beta$-function.

\emph{Acknowledgements}:
C.K wishes to thank E.G.Floratos for discussions. The work of D.V.N. is
supported by DOE grant DEFG03-95-Er-40917.





\end{document}